\documentclass{Interspeech}

\interspeechcameraready

\usepackage{tikz}
\usetikzlibrary{shapes.geometric}

\title{Effective Context in Neural Speech Models}

\author[affiliation={}]{Yen}{Meng}
\author[affiliation={}]{Sharon}{Goldwater}
\author[affiliation={}]{Hao}{Tang}

\affiliation[nocounter]{The Centre for Speech Technology Research}{University of Edinburgh}{United Kingdom}
\email{yen.meng@ed.ac.uk, sgwater@inf.ed.ac.uk, hao.tang@ed.ac.uk}

\usepackage{comment}
\usepackage{microtype} 
\usepackage{cite} 
\usepackage[textsize=footnotesize]{todonotes}

\newcommand{\wsjtrain}{{\footnotesize\texttt{si284}}}
\newcommand{\wsjdev}{{\footnotesize\texttt{dev93}}}

\newcommand{\lstest}{{\footnotesize\texttt{test-clean}}}

\begin{document}

\maketitle

\begin{abstract}
Modern neural speech models benefit from having longer context, and many approaches have been proposed to increase the maximum context a model can use. However, few have attempted to measure how much context these models actually use, i.e., the effective context. Here, we propose two approaches to measuring the effective context, and use them to analyze different speech Transformers. For supervised models, we find that the effective context correlates well with the nature of the task, with fundamental frequency tracking, phone classification, and word classification requiring increasing amounts of effective context. For self-supervised models, we find that effective context increases mainly in the early layers, and remains relatively short---similar to the supervised phone model. Given that these models do not use a long context during prediction, we show that HuBERT can be run in streaming mode without modification to the architecture and without further fine-tuning.

\end{abstract}

\section{Introduction}
\label{sec:intro}
The recent success of speech and language processing systems can be largely attributed to better modeling of context.
Various speech tasks benefit from contextualized speech embeddings learned with self-supervision~\cite{apc, cpc, wav2vec2, hubert, wavlm}. Automatic speech recognition, summarization, question answering, and language modeling have all been shown to improve when models are engineered to have a wider context~\cite{hori20_interspeech, large_context_asr, flynn2023much, transformerxl, infinite-context, mohtashami2023landmark}.
However, observing an improved performance does not necessarily imply that a wider context is actually used.
Failing to properly attribute where the improvement is from gives the false impression that longer context is always better.
It is possible that a wider context is not even necessary to achieve competitive performance.

In this work, we distinguish between the context a model has access to (i.e., the context by design) and the context that a model actually uses (the effective context).
The amount of context available to the model can be increased by changing the model architecture~\cite{transformerxl, infinite-context, mohtashami2023landmark}.
However, the context that the model actually uses will be affected by various other design choices, such as the data and the training algorithm.
LSTMs, for example, have access to the entire history but also have difficulty making full use of it~\cite{tbptt-rnn}.
Several studies have presented indirect evidence that Transformers, despite having access to the entire input sequence, do not make use of the full context.
They approach this problem by modifying the model architecture~\cite{usefulness_attention, branchformer, parcollet2023sumformer} or by analyzing attention maps~\cite{usefulness_attention, shim2022understanding}.
A few other studies~\cite{pasad2021layer, pasad2023comparative, pasad2023self, de_seyssel_probing_2022, choi2024self}, though not directly studying the amount of context, have probing results that suggest the limited use of context in Transformers.
Overall, these approaches are ad-hoc and limited either to a particular model architecture or to a particular task.

In this work, we approach the problem of measuring effective context from first principles.
If a change to a part of the input does not affect the output much, then that part of the input is effectively not used by the model and hence not in the effective context.
We propose two approaches, one based on truncation of the input and the other based on the Jacobian with respect to the input.
The truncation approach is more intuitive---truncating input frames not in the effective context should hardly affect the output.
The Jacobian approach (similar to gradients) measures the effect of infinitesimal changes on the input.
Both approaches are model-agnostic, not tied to specific loss functions, do not require labels, and can be applied to any layer of a network.

We apply the proposed approaches to measure effective context of speech Transformers, as they are the dominant architecture in both supervised and self-supervised settings.
In the supervised setting, we find that the amount of effect context used in Transformers for predicting f0, phones, and words increases in that order. 
In the self-supervised setting, we find that the effective context of self-supervised Transformers, specifically wav2vec 2.0 \cite{wav2vec2}, HuBERT \cite{hubert}, and WavLM \cite{wavlm}, is shorter than that for predicting phones in the supervised setting.
We also observe a clear correlation between the amount of effective context and probing performance on phones and words.

Given the relatively short effective context of self-supervised Transformers, limiting the history and lookahead of these models should not hurt their performance much.
This implies that we can readily run pretrained Transformers in low-latency (400ms) streaming mode without any modification to the architecture or further training.
Using a simple probing classifier, our experiments show that, compared to a phone error rate of 11.9\% when using full utterances, we only observe a 0.6\% degradation with the full history, and a 1.5\% degradation with a 2s history.

\section{Measuring Effective Context}

We define effective context based on the following principle.
For an input utterance of $T$ frames $x_1, \dots, x_T$, we are interested in how a function $f$ changes as we change the frames.
At a high level, if changes made to frame $x_t$ do not change the output of $f$ by much, then $x_t$ is not part of the effective context of $f$.
The function $f$ can be 
the computation from the input to any of the intermediate layers, or all the way to the loss.
The output of $f$ typically has multiple frames, i.e., $h_1, \dots, h_T = f(x_1, \dots, x_T)$, and we will use $[f(x_1, \dots, x_T)]_t$ to denote $h_t$.
Below we detail two changes that can made to the frames, leading to the truncation approach and the Jacobian approach.

\subsection{The Truncation Approach}
\label{sec:truncation-setup}

If we are interested in whether a frame $x_t$ is in the effective context or not, the most intuitive approach is to remove $x_t$ from the input to $f$.
Removing single frames is not very meaningful as, in speech, a lot can be inferred from neighboring frames.
Instead, for models that can make use of the full context, we make the assumption (which we will later verify) that $f$ has a symmetric effective context. 
We choose a window of size $2W+1$ frames centered at $t$ and truncate the frames outside of the window.
We then compute
\begin{align}
d\Bigg( \Big[f(x_1, \dots, x_T)\Big]_t, \Big[f(x_{t-W},  \dots, x_{t+W})\Big]_t \Bigg)
\end{align}
for $t=1, \dots, T$ of an utterance and average over a data set, where $d$ is a distance function that can be the $\ell_2$ distance or a performance metric, such as phone probing error rate.
The truncation approach has been inspired by similar ideas used to study the context of text-based language models~\cite{khandelwal2018sharp, o2021context, levy2024same}.

\begin{figure}
    \centering
    \includegraphics[height=7.6em]{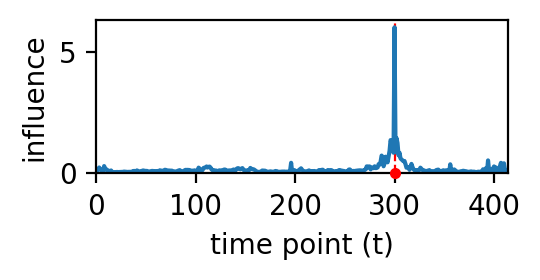}
    \includegraphics[height=7.6em]{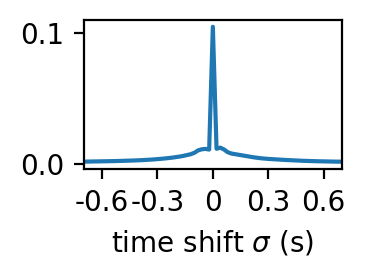}
    \vspace{-2mm}
    \caption{
    Examples illustrating influence. Left: The x-axis is the time point $\tau$ of the input utterance $x$, and the y-axis is the calculated \textbf{influence} value $s(t, \tau)$ for the timepoint $t=300$ (red dot). Right: The \textbf{relative influence} $S(\sigma)$ (normalized), where the x-axis is the time shift to the center frame.
    }.
    \vspace{-2mm}
    \label{fig:mp-influence-example}
\end{figure}

\subsection{The Jacobian Approach}
\label{ref:influence-setup}

Instead of explicitly truncating, another approach is to make changes to the input.
Formally, we want to compute
\begin{align}
\Bigg\lVert \Big[f(\dots, x_\tau, \dots)\Big]_t - \Big[f(\dots, x_\tau + \epsilon, \dots)\Big]_t \Bigg\rVert_\text{F}
\end{align}
where $\epsilon$ is the change made to the frame $x_\tau$ $\in \mathbb{R}^K$.
When $\epsilon$ is infinitesimal, the above term becomes the Frobenius norm of the Jacobian (or the $\ell_2$ norm of the gradient when $f$ is $\mathbb{R}^K \to \mathbb{R}$).
We call this quantity, denoted as $s(t, \tau)$, the \textbf{influence}
of $x_\tau$ on $h_t$ (or $[f(x_1, \dots, x_T)]_t$). See Fig.~\ref{fig:mp-influence-example} (left) for an example.
Note that, compared to the truncation approach, $t$ and $\tau$ are decoupled here, and no windowing or symmetry assumptions are made. 

In practice, most automatic differentiation packages do not explicitly calculate the full Jacobian matrix but rather the vector-Jacobian product.
To compute the full Jacobian matrix of a hidden vector $h_t \in \mathbb{R}^D$, we simply compute the vector-Jacobian product on the $D$-dimensional standard basis, backpropagating $D$ times.
In other words, the influence is computed as 
\begin{align}
  s(t, \tau) = \sqrt{\sum_{k=1}^K \sum_{d=1}^D
  \left( e_d^\top \frac{\partial h_t}{\partial x_{\tau, k}} \right)^2}
  \label{eq:influence}
\end{align}
where $e_d$ is $d$-th element of the $D$-dimensional standard basis, and $\frac{\partial h_t}{\partial x_{\tau}}$ is the Jacobian matrix of time $t$ with respect to the input at time $\tau$.

Similar to the truncation approach, we introduce windowing and study the influence relative to the center frame.
For a data set of $N$ utterances, each of which has $T_n$ frames with the measured influence $s_n(t, \tau)$ of $f$, for $n=1, \dots, N$, the \textbf{relative influence} at time shift $\sigma$ is then defined as
\begin{align}
    S(\sigma) = \sum_{n=1}^{N} \sum_{t=1}^{T_n} s_n(t, t + \sigma).
    \label{eq:normsums}
\end{align}
For example, when $S(-w)$ is high, the influence on hidden vectors from the input that is $w$ frames before is, on average, high.
Given a window size of $2W+1$ frames, we compute the relative influence for time shifts $\sigma = -W, -W+1, \dots, W$.
The Jacobian matrices are not scale invariant, so we normalize the relative influence such that $\sum_{\sigma=-W}^W S(\sigma) = 1$. See Fig.~\ref{fig:mp-influence-example} (right) for an example. After normalization, the relative influence of different models of layers can be compared. In the relative influence plots, we convert frames into seconds for easier reference.

To contrast with other approaches, saliency~\cite{simonyan2013deep} of certain parts of the input used to explain model prediction is typically computed by taking the gradient to the input.
Though the mathematical formulation (Eq.~\ref{eq:influence}) of the Jacobian approach happens to be a generalization of saliency, our goal is different---to estimate the effective context rather than explain particular outputs---and the generalization allows us to apply the approach to any layer (not just the final output).
Other explainability methods, such as LIME~\cite{LIME} and SHAP~\cite{SHAP}, are not suitable for our purpose, as they do not necessarily inform us about the effective context, are not task-agnostic, or cannot be applied to intermediate layers.

\begin{figure*}
    \centering
    \definecolor{f0}{HTML}{66cccc}
    \definecolor{phn}{HTML}{cc66cc}
    \definecolor{wrd}{HTML}{cccc66}
    \definecolor{wavlm}{HTML}{1f77b4}
    \definecolor{w2v2}{HTML}{ff7f0e}
    \definecolor{hubert}{HTML}{2ca02c}
    \definecolor{L1}{HTML}{98d594}
    \definecolor{L4}{HTML}{4bb062}
    \definecolor{L8}{HTML}{157f3b}
    \definecolor{L12}{HTML}{00441b}
    \begin{tikzpicture}
    \draw[f0, ultra thick] (-14.6, -0.4) -- (-14.0, -0.4);
    \node[right] at (-14.0, -0.4) {f0};
    \draw[phn, ultra thick] (-13.2, -0.4) -- (-12.6, -0.4);
    \node[right] at (-12.6, -0.4) {phone};
    \draw[wrd, ultra thick] (-11.4, -0.4) -- (-10.8, -0.4);
    \node[right] at (-10.8, -0.4) {word};
    \node at (-12.5, -2.4) {\includegraphics[height=3.5cm]{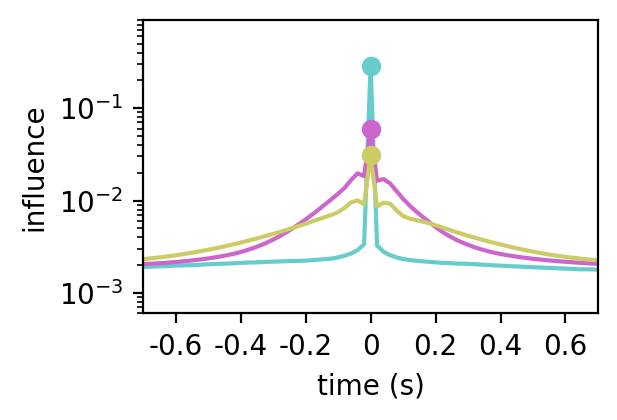}};
\node at (-14.55, -3.8) {(a)};

    \draw[black!33, ultra thick, dashed] (-6.8, 0.4) -- (-6.2, 0.4);
    \node[right] at (-6.2, 0.4) {untrained L9};
    \draw[L1, ultra thick] (-9.4, 0) -- (-8.8, 0);
    \node[right] at (-8.8, 0) {HuBERT L1};
    \draw[L4, ultra thick] (-6.8, 0) -- (-6.2, 0);
    \node[right] at (-6.2, 0) {HuBERT L4};
    \draw[L8, ultra thick] (-9.4, -0.4) -- (-8.8, -0.4);
    \node[right] at (-8.8, -0.4) {HuBERT L8};
    \draw[L12, ultra thick] (-6.8, -0.4) -- (-6.2, -0.4);
    \node[right] at (-6.2, -0.4) {HuBERT L12};    
    \node at (-7.2, -2.4) {\includegraphics[height=3.5cm]{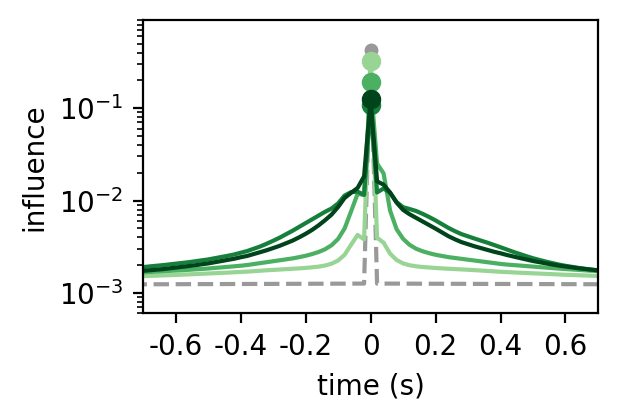}};
    \node at (-9.3, -3.8) {(b)};

    \draw[ultra thick, w2v2] (-3.6, 0) -- (-2.8, 0);
    \node[regular polygon, regular polygon sides=3, fill=w2v2, inner sep=1.2pt] at (-3.2, 0) {};
    \node[right] at (-2.8, 0) {wav2vec 2.0};
    \draw[ultra thick, hubert] (-0.8, 0) -- (-0.0, 0);
    \node[regular polygon, regular polygon sides=4, rotate=45, fill=hubert, inner sep=1.8pt] at (-0.4, 0) {};
    \node[right] at (-0.0, 0) {HuBERT};
    \draw[ultra thick, wavlm] (-3.6, -0.4) -- (-2.8, -0.4);
    \node[circle, fill=wavlm, inner sep=1.8pt] at (-3.2, -0.4) {};
    \node[right] at (-2.8, -0.4) {WavLM};
    \draw[ultra thick, black!33] (-0.8, -0.4) -- (-0.0, -0.4);
    \node[regular polygon, regular polygon sides=3, rotate=180, fill=black!33, inner sep=1.4pt] at (-0.4, -0.4) {};
    \node[right] at (-0.0, -0.4) {untrained};

    \node at (-1.6, -2.4) {\includegraphics[height=3.5cm]{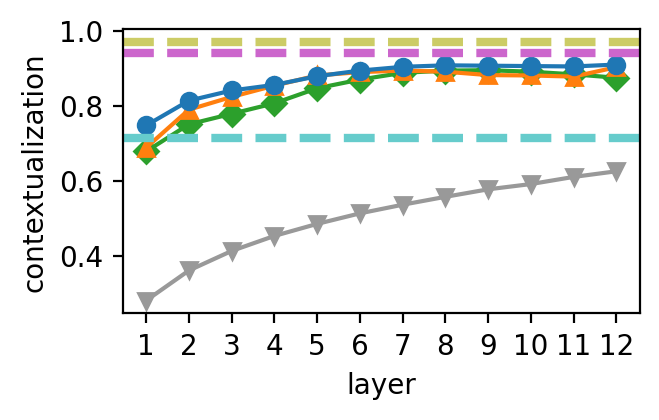}};
    \node[right, wrd] at (1.0, -0.9) {word};
    \node[right, phn] at (1.0, -1.2) {phone};
    \node[right, f0] at (1.0, -1.8) {f0};
    \node at (-3.8, -3.8) {(c)};
    \end{tikzpicture}
    \vspace{-6mm}
    \caption{
    (a) Relative influence in the final layers of supervised 6-layer Transformer models trained for different tasks. The y-axis is on a log scale and the x-axis is only shown between $\pm0.7s$, although the relative influence values were computed with a window size of 5 seconds on both sides. Dots show the heights of the center peaks. (b) As in (a) but for different layers of HuBERT. 
    (c) Contextualization of different models and layers. 
    The horizontal lines are values of the supervised models on the final layer.}
    \vspace{-2mm}
    \label{fig:influence-align}
\end{figure*}
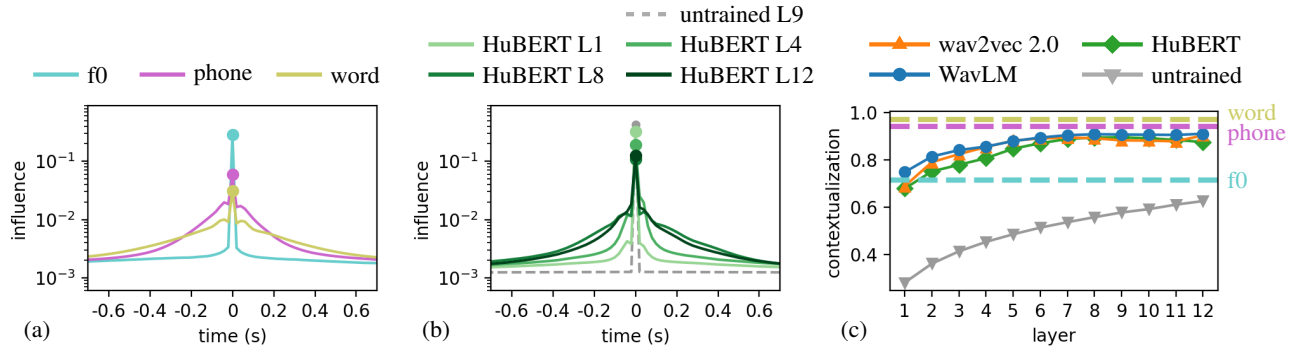

\begin{figure}
    \centering
    \definecolor{c1}{HTML}{b7d5ea}
    \definecolor{c2}{HTML}{8dc1de}
    \definecolor{c3}{HTML}{60a7d2}
    \definecolor{c4}{HTML}{3b8bc3}
    \definecolor{c5}{HTML}{1d6cb1}
    \definecolor{c6}{HTML}{084e98}
    \definecolor{c7}{HTML}{08306b}
    \begin{tikzpicture}
    \draw[c1, ultra thick] (0, 0) -- (0.8, 0);
    \node[circle, fill=c1, inner sep=1.2pt] at (0.4, 0) {};
    \node[right] at (0.8, 0) {0.0s};
    \node[circle, fill=c2, inner sep=1.2pt] at (2.2, 0) {};
    \draw[c2, ultra thick] (1.8, 0) -- (2.6, 0);
    \node[right] at (2.6, 0) {0.2s};
    \draw[c3, ultra thick] (3.6, 0) -- (4.4, 0);
    \node[circle, fill=c3, inner sep=1.2pt] at (4.0, 0) {};
    \node[right] at (4.4, 0) {0.4s};
    \draw[c4, ultra thick] (5.4, 0) -- (6.2, 0);
    \node[circle, fill=c4, inner sep=1.2pt] at (5.8, 0) {};
    \node[right] at (6.2, 0) {0.8s};
    \draw[c5, ultra thick] (0, -0.4) -- (0.8, -0.4);
    \node[circle, fill=c5, inner sep=1.2pt] at (0.4, -0.4) {};
    \node[right] at (0.8, -0.4) {2.0s};
    \node[circle, fill=c6, inner sep=1.2pt] at (2.2, -0.4) {};
    \draw[c6, ultra thick] (1.8, -0.4) -- (2.6, -0.4);
    \node[right] at (2.6, -0.4) {4.0s};
    \draw[c7, ultra thick] (3.6, -0.4) -- (4.4, -0.4);
    \node[circle, fill=c7, inner sep=1.2pt] at (4.0, -0.4) {};
    \node[right] at (4.4, -0.4) {8.0s};
    \draw[black, ultra thick] (5.4, -0.4) -- (6.2, -0.4);
    \node[circle, fill=black, inner sep=1.2pt] at (5.8, -0.4) {};
    \node[right] at (6.2, -0.4) {$\infty$};
    \end{tikzpicture}
    \\
    \includegraphics[width=3.8cm]{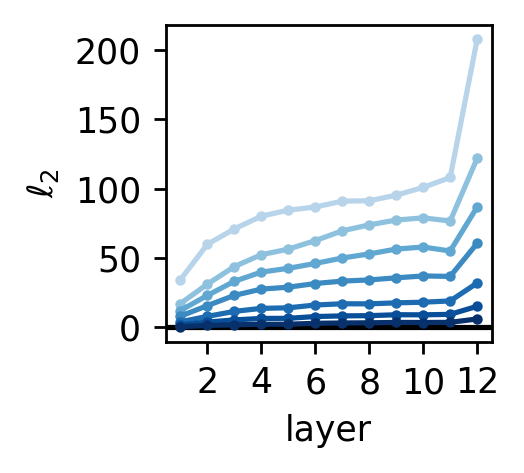}
    \includegraphics[width=3.8cm]{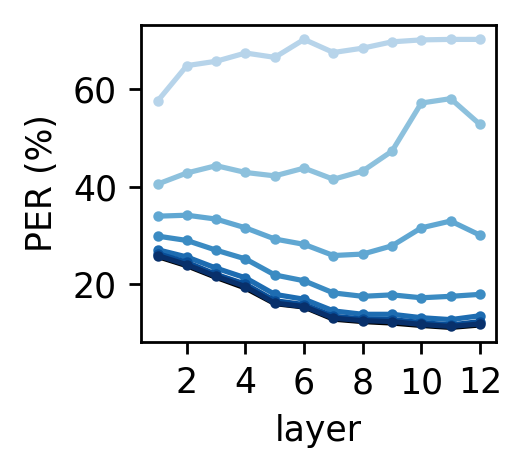}
    \vspace{-2mm}
    \caption{The change of output in terms of the $\ell_2$ distance (left) and the phone error rates (right), as we vary the window size of input to HuBERT (different coloured lines).}
    \vspace{-2mm}
    \label{fig:symmetric-window-hubert}
\end{figure}
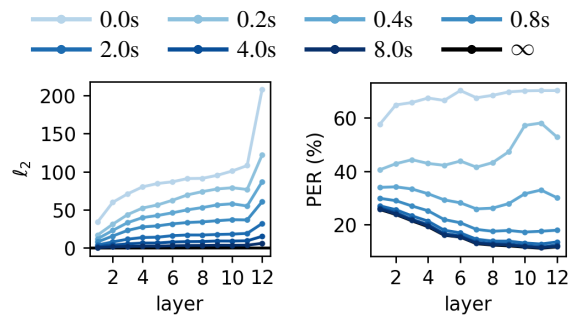

\section{Experiments}

Since there isn't any prior work on measuring effective context in speech models, our goal is to provide a foundation for future work to measure the effective context for a broader range of models and architectures. In this section, we select a few models to analyze and showcase the utility of our approach. 

\subsection{Pilot experiments with truncation}
\label{sec:truncation-symmetric}

For the first set of experiments, we measure the effective context of a pretrained HuBERT (base)\footnote{The average utterance length of LibriSpeech where HuBERT is trained on is 12.3s. Each audio example is cropped to 15.6s during pretraining.}.
We follow~\cite{apc, cpc, apc-journal} and train a linear probing classifier (using multinomial logistic regression) on WSJ to predict phones for each layer of HuBERT. 
We then measure how the phone error rate changes as we truncate the frames and only keep a window of frames.
This measures how much change in the hidden vectors is needed to flip the label of a classifier.
We also directly measure the $\ell_2$ distance between the representations computed from the original and truncated inputs.

Results on \wsjdev\ are shown in Fig.~\ref{fig:symmetric-window-hubert}.
Although very narrow truncation windows have a large effect, by the time the window size reaches 2.0s, PER is similar for truncated and un-truncated input across all layers. The same is true for $\ell_2$ with a window size of 4.0s. 
We conclude that the effective context of HuBERT is about 2.0s on either side of the center frame 
(and as a reference the average utterance length in \wsjdev\ is 7.8s).

Based on the trend in $\ell_2$ distance, it seems that the effective context increases after each layer, though the trend is less clear in the PER case.
It is relatively difficult to summarize the layerwise results with the truncation approach, but these initial results will be useful later as we check for consistency.

\subsection{Results with the Jacobian approach}
\label{sec:influence-results}

We follow the same setting as the truncation approach.
However, instead of immediately comparing results on HuBERT layers, we first analyze a set of supervised Transformers for predicting f0, phones, and words.
We will then use these results to contextualize the Jacobian results on HuBERT layers, and further extend them to wav2vec 2.0 and WavLM.
We compute the relative influence with a 10s window, sufficiently large given that the $\ell_2$ does not change much already within a 4.0s window.

\vspace{0.5em}
\noindent\textbf{Effective context of supervised models}\hspace{0.25em}
For all supervised models, we use the same 6-layer Transformer with sinusoidal position encoding. Training is done on the \wsjtrain\ subset of WSJ.
The input to the Transformers is 80 dimensional, consisting of 40-dimensional Mel spectrograms stacked every two frames.
The frame rate is 20ms, matching those of HuBERT, wav2vec 2.0, and WavLM.
For $f_0$ tracking, the targets are extracted with the PYIN algorithm~\cite{pyin}.
Phone and word classification are conducted at the frame level, following \cite{tbptt-rnn}, with labels obtained through forced alignments computed with an HMM-GMM from Kaldi.
For reference, these models achieve 8.2 Hz RMSE on $f_0$ tracking, a 12.4\% error rate on phone classification, and a 26.2\% error rate on word classification on \wsjdev.

In Fig.~\ref{fig:influence-align}(a), we show the relative influence of the final layer of the 6-layer Transformer, as we are interested in the effective context needed for the tasks.
First, the relative influence in all three tasks is roughly symmetric and drops quickly as we move away from the center.
This suggests that most of the influence comes from the center frame (and showing the range within $\pm$0.7s in Fig.~\ref{fig:influence-align} is already sufficient).
Second, predicting f0 uses the least amount of effective context, followed by predicting phones.
Predicting words uses the largest effective context, as can be seen from the lower center peak and the fatter tail.

\vspace{0.5em}
\noindent\textbf{Effective context of self-supervised models}\hspace{0.25em}
For the self-supervised setting, we focus on masked prediction models, HuBERT, wav2vec2.0, and WavLM, which are 12-layer Transformers pretrained on the 960-hour subset of LibriSpeech (base models). These models take wave samples as input, so we measure the norms at the output of the convolution feature extractor (i.e., the input before the Transformer encoder) so that the context measurement for all models only includes Transformer layers.

Fig.~\ref{fig:influence-align}(b) shows the relative influence of layer 1, 4, 8, 12, compared to a late layer in an untrained Transformer.
We again see that the relative influence for all layers decays sharply moving away from the center and the shape is roughly symmetric.
The effective context seems to increase through the layers of the pretrained model, but remains narrow for the untrained model.

We then compare the HuBERT layers in Fig.~\ref{fig:influence-align}(b) with the supervised results in Fig.~\ref{fig:influence-align}(a).
The effective context of higher layers of HuBERT is close to the 6-layer phone model but much smaller compared to the word model. 
This might explain a recent finding that self-supervised speech representations are more phonetic than semantic \cite{choi2024self}.
The results presented above were on WSJ \wsjdev, but we also analyzed relative influence for HuBERT on  LibriSpeech \lstest (average utterance length of 7.4s), with very similar results.

\vspace{0.5em}
\noindent\textbf{Layerwise analysis}\hspace{0.25em}
Though relative influence provides a detailed shape of the effective context, it is difficult to compare quantitatively among layers and across models.
For this, it is desirable to summarize the relative influence curve using a single value. While there are various options (such as reporting the width that contains some threshold proportion of the mass), the values depend both on the choice of threshold and on the window size used to compute the relative influence. 

In practice, we find that a very simple statistic works well: we use $1 - S(0)$, where $S(0)$ is the relative influence at the center frame, and refer to this as the \textbf{contextualization}. It is 0 when only the immediate input frame influences the result and approaches 1 if there is almost no influence from the center frame.
The exact values still depend on the window size used to compute the relative influence, but we can compare relative values of contextualization for different models/layers, as long as all values are based on the same window size (as here).\footnote{We found that qualitative trends for contextualization depended less on window size than other statistics we considered early on.}

Fig.~\ref{fig:influence-align}(c) shows the contextualization of wav2vec 2.0, HuBERT, WavLM, and an untrained Transformer (randomly initialized following~\cite{fairseq}).
The contextualization of pretrained models increases from layer 1 to layer 7, then plateaus.
When we include the contextualization of the 6-layer phone and word predictor, it is now clear that the pretrained models are less contextualized than the phone predictor.
Interestingly, the effective context of the untrained model increases throughout the layers.
This potentially explains the increase in performance throughout the layers for speaker tasks in untrained Transformers \cite{mohamed2024orthogonality}.
Comparing the pretrained models and the untrained one, it is clear that self-supervised training is a process of learning contextualization.

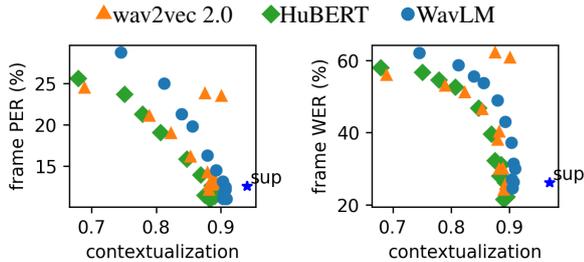
\begin{figure}[t] 
    \centering
    \definecolor{wavlm}{HTML}{1f77b4}
    \definecolor{w2v2}{HTML}{ff7f0e}
    \definecolor{hubert}{HTML}{2ca02c}
    \begin{tikzpicture}
    \node[regular polygon, regular polygon sides=3, fill=w2v2, inner sep=1.3] at (-2.0, 0) {};
    \node[right] at (-2.0, 0) {wav2vec 2.0};
    \node[regular polygon, regular polygon sides=4, rotate=45, fill=hubert, inner sep=1.8] at (0.2, 0) {};
    \node[right] at (0.2, 0) {HuBERT};
    \node[circle, fill=wavlm, inner sep=1.8] at (2.0, 0) {};
    \node[right] at (2.0, 0) {WavLM};
    \end{tikzpicture}
    \includegraphics[width=3.9cm]{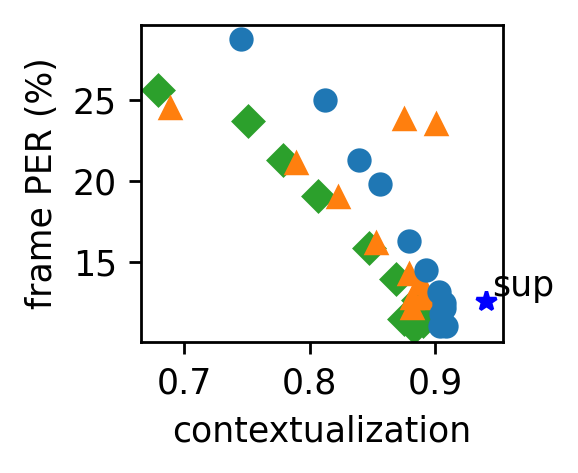}  
    \includegraphics[width=3.9cm]{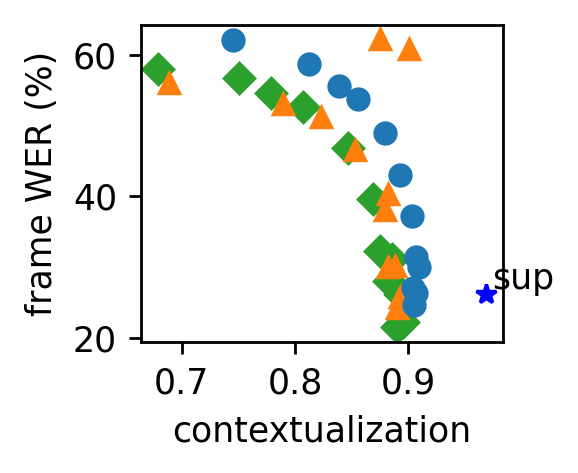}
    \vspace{-2mm}
    \caption{Relation between contextualization and probing performance on phones (left) and words (right). Each point represents a layer of a model. The last layer of the supervised 6-layer Transformer is annotated.}
    \vspace{-2em}
    \label{fig:context-probing}
\end{figure}

Fig.~\ref{fig:context-probing} illustrates the relation between phone and word probing (again using multinomial logistic regression classifiers) and contextualization. We observe a clear correlation between lower error rates and greater contextualization. The only exceptions to this general pattern are the last two layers of the wav2vec 2.0 model (the two outlier points in the plots), which other studies have also found to behave idiosyncratically~\cite{pasad2021layer,pasad2023comparative,mohamed2024orthogonality}. 
Fig.~\ref{fig:context-probing} suggests that higher contextualization does not guarantee higher probing accuracy; rather,
there seems to be a minimum amount of effective context that a model needs to achieve in order to predict phones and words well. The phone error rate decreases steadily starting from the first layer, whereas for predicting words, the probing error rate remains high in the first few layers, and starts to drop once the model has reached a certain degree of contextualization. 

\section{Simulating a Streaming HuBERT}
\label{sec:applications}

Given that the effective context of pretrained Transformers is not long, we should be able to truncate their context and run them in a low-latency streaming mode without much performance loss.
This would give us a streaming representation extractor.
In contrast to other streaming models~\cite{dual-mode-asr, cai2023efficient, doutre2021improving, kumar2024xlsr}, this approach does not require modifications to the architecture or attention and does not require further training.
We take HuBERT as an example and run it as a sliding window over the input frames.
For streaming, the choice to make is the amount of history and the amount of lookahead to use.
The history largely determines the overall memory usage, while the lookahead determines the latency.
The window of input to HuBERT is the concatenation of the history and the lookahead.

\begin{figure}[t]
    \centering
    \definecolor{infhist}{HTML}{1f77b4}
    \definecolor{lookahead1}{HTML}{ffa3a3}
    \definecolor{lookahead2}{HTML}{ff5555}
    \definecolor{lookahead3}{HTML}{db0202}
    \begin{tikzpicture}
    \draw[lookahead1, ultra thick] (-3.4, -0.4) -- (-2.6, -0.4);
    \node[fill=lookahead1, regular polygon, regular polygon sides=3, rotate=90, inner sep=1.2pt] at (-3.0, -0.4) {};
    \node[right, align=center] at (-2.6, -0.4) {0ms lookahead};
    \node[fill=lookahead3, regular polygon, regular polygon sides=3, rotate=90, inner sep=1.2pt] at (0.4, -0.4) {};
    \draw[lookahead3, ultra thick] (0.0, -0.4) -- (0.8, -0.4);
    \node[right, align=center] at (0.8, -0.4) {400ms lookahead};
    \draw[lookahead2, ultra thick] (-3.4, -0.8) -- (-2.6, -0.8);
    \node[fill=lookahead2, regular polygon, regular polygon sides=3, rotate=90, inner sep=1.2pt] at (-3.0, -0.8) {};
    \node[right, align=center] at (-2.6, -0.8) {200ms lookahead};
    \draw[infhist, ultra thick] (0.0, -0.8) -- (0.8, -0.8);
    \node[fill=infhist, regular polygon, regular polygon sides=3, rotate=-90, inner sep=1.2pt] at (0.4, -0.8) {};
    \node[right] at (0.8, -0.8) {unlimited history};
    \end{tikzpicture}
    \\
    \includegraphics[width=3.9cm]{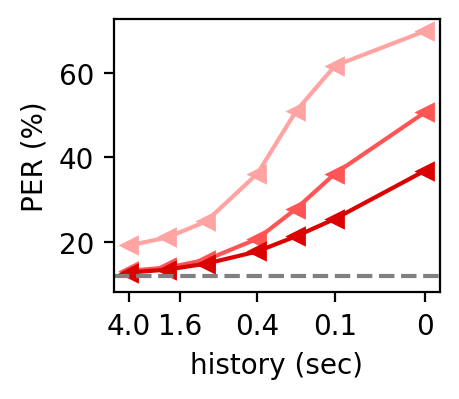}
    \includegraphics[width=3.9cm]{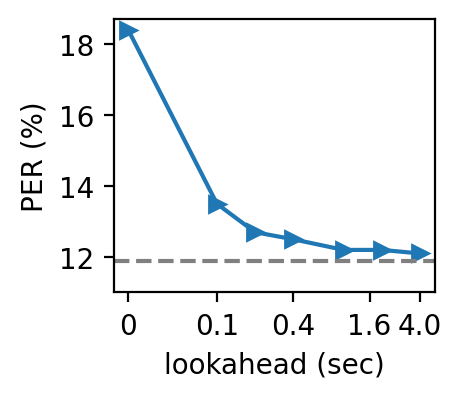}
    \vspace{-2mm}
    \caption{Results of different streaming settings. The dashed line represents the probing error rate with full context. We vary the number of lookahead with unlimited history (right), and vary the number of history with different amounts of lookahead (left).
    }
    \vspace{-2em}
    \label{fig:streaming-setting}
\end{figure}
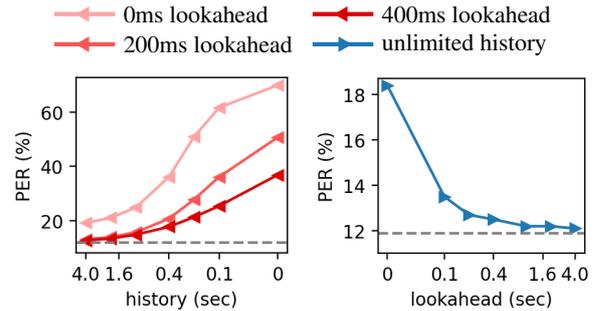

Fig.~\ref{fig:streaming-setting} shows the results of phone probing on layer 9 of a streaming HuBERT, using the probing classifier from \S\ref{sec:truncation-symmetric} without retraining.
For an unlimited history (right panel of Fig.~\ref{fig:streaming-setting}), a lookahead of 400ms is sufficient to recover most of the phone error rates, achieving 12.5\% compared to 11.9\% when using the full context.
In terms of the history to truncate if we fix the lookahead to 400ms (left panel of Fig.~\ref{fig:streaming-setting}, lowest curve), a history of 2s is sufficiently strong, achieving 13.7\%.
The results suggest that HuBERT representations can be extracted using history and lookahead values within the range used for recent streaming ASR systems~\cite{dual-mode-asr, cai2023efficient, doutre2021improving, kumar2024xlsr}, with little loss of fidelity.

\section{Discussion and Conclusion}
In this work, we designed two complementary approaches to measuring effective context from first principles.
The truncation approach is a direct modification to the input and can be easily verified by task performance difference, but how and how much we can truncate requires empirical verification.
The Jacobian approach is less clear in terms of task performance as changes to the input are not actually made. However, it does not require any assumptions regarding the shape of the relative influence curve, and allows us to examine these directly, 
providing a more fine-grained view of the influence of context at different distances from the current frame. 
Both approaches are easy to implement, trivially parallelizable, and can be applied regardless of architecture or training objective.

By combining these two approaches, we found strong evidence that effective context differs for different supervised tasks using the same model architecture, and that self-supervised Transformers have relatively short effective context. Our results also indicate that for pretrained models, there is a strong correlation between contextualization and probing performance.
Nevertheless, contextualization alone cannot predict performance.
In particular, our 6-layer supervised models have a larger effective context than the 12-layer pretrained models, but have slightly worse phone and word error rates. This implies that other factors, such as the structure of the representations or the amount of training data, are also important.
Our work, however, already provides us with sufficient insights into how self-supervised models use context and to have practical applications.
We are able to simulate a streaming HuBERT with low latency and minor performance degradation.
In future work, we hope to use the tools developed here to further examine effective context in other types of models, such as ECAPA-TDNN~\cite{ecapa} or Whisper~\cite{whisper}.
Designing models that have large context windows does not necessarily contribute to solving the long-context modeling problem unless models actually use that context, and we hope that this work offers an opportunity to design long-context models from first principles.

\clearpage
\bibliographystyle{IEEEtran}
\bibliography{mybib}

\end{document}